\documentclass[acmsmall,screen,nonacm]{acmart}
\setcopyright{none}
\AtBeginDocument{%
  }

\usepackage[svgnames, table]{xcolor}
\usepackage{tcolorbox}
\usepackage{algpseudocode}
\tcbuselibrary{skins, breakable}
\usepackage{graphicx}
\usepackage{cleveref}
\usepackage{amsmath}
\usepackage{subcaption}
\usepackage{booktabs}
\usepackage{multirow}
\usepackage{pgfplots}
\usepackage{varwidth}
\pgfplotsset{compat=1.18}
\usepackage{caption}
\usepackage{soul}
\usepackage{algorithm}
\usepackage{algpseudocode}

\definecolor{repColor}{HTML}{E3F2FD}     
\definecolor{maint1Color}{HTML}{FFEBEE}  
\definecolor{maint2Color}{HTML}{FFFDE7}  
\definecolor{maint3Color}{HTML}{E8F5E9}  
\definecolor{borderColor}{HTML}{E0E0E0}  

\definecolor{repoBrave}{HTML}{0072B2}
\definecolor{repoHomeAssistant}{HTML}{E69F00}
\definecolor{specificityGeneric}{HTML}{009E73}
\definecolor{specificityDetailed}{HTML}{CC79A7}

\tcbset{chatbubble/.style={
    enhanced,
    breakable,
    arc=5mm,         
    outer arc=5mm,
    boxrule=1pt,     
    colframe=borderColor,
    boxsep=5pt,
    left=8pt, right=8pt, top=2pt, bottom=2pt, 
    drop shadow=gray!20, 
    fontupper=\scriptsize, 
    fonttitle=\sffamily\bfseries\scriptsize, 
    coltitle=black!60, 
    boxed title style={empty, boxsep=3pt}, 
}}

\newcommand{\chatLeft}[3]{%
    \begin{tcolorbox}[chatbubble,
        colback=#2,
        title={#1},
        attach boxed title to top left={yshift=-2mm, xshift=5mm}, 
        hbox,
        unbreakable,
        flush left, 
        before=\par\vspace{4pt}\noindent, 
        after=\par
        ]
    \begin{varwidth}{0.9\columnwidth}
    #3
    \end{varwidth}
    \end{tcolorbox}
}

\newcommand{\chatRight}[3]{%
    \begin{tcolorbox}[chatbubble,
        colback=#2,
        title={#1},
        halign title=right, 
        attach boxed title to top right={yshift=-2mm, xshift=-5mm}, 
        hbox, 
        unbreakable,
        flush right, 
        before=\par\vspace{4pt}\noindent, 
        after=\par
        ]
    \begin{varwidth}{0.9\columnwidth}
    #3
    \end{varwidth}
    \end{tcolorbox}
}

\newcommand{\tool}{BRQGen}
\newcommand{\positiveDataset}{$\mathcal{B}^{+}$}
\newcommand{\negativeDataset}{$\mathcal{B}^{-}$}

\newenvironment{DIFnomarkup}{\ignorespaces}{\unskip}
\begin{document}

\title{\tool: Towards Adaptive Completeness Assessment and Follow-Up Question Generation for Bug Reports}

\author{Tevfik Emre Sungur}
\correspondingauthor
\authornote{Both authors contributed equally to this research.}
\email{tevfikemre.sungur@tum.de}
\orcid{0009-0003-7140-7131}
\affiliation{%
  \institution{Technical University of Munich}
  \department{Department of Computer Science, TUM School of Computation, Information and Technology}
  \city{Munich}
  \country{Germany}
}

\author{Akif Emre Köşüş}
\authornotemark[1]
\email{emrekosus@bilkent.edu.tr}
\orcid{0009-0000-4800-1326}
\affiliation{%
  \institution{Bilkent University}
  \department{Department of Electrical and Electronics Engineering}
  \city{Ankara}
  \country{Türkiye}
}

\author{Uğurcan Doğruçam}
\email{ugurcan.dogrucam@ug.bilkent.edu.tr}
\orcid{0009-0005-7636-9546}
\affiliation{%
  \institution{Bilkent University}
  \department{Department of Computer Engineering}
  \city{Ankara}
  \country{Türkiye}
}

\author{Eray Tüzün}
\email{eraytuzun@cs.bilkent.edu.tr}
\orcid{0000-0002-5550-7816}
\affiliation{%
  \institution{Bilkent University}
  \department{Department of Computer Engineering}
  \city{Ankara}
  \country{Türkiye}
}

\renewcommand{\shortauthors}{Sungur et al.}

\begin{abstract}
  \textbf{Context:} Software maintenance relies heavily on bug reports provided by end-users and testers. However, a large portion of these reports lack the necessary information required for developers to reproduce issues. This information gap increases triage time and often leads to valid bugs being closed as \textit{non-reproducible}. While early studies utilized large language models (LLMs) to enhance low-quality reports, these approaches introduce incorrect information when the initial input lacks sufficient semantic content. More recently, LLMs have been employed to generate context-aware follow-up questions. However, existing methods rigidly ask a fixed number of questions regardless of whether the report is actually incomplete. Hence, it is essential to decide on the completeness of the report and ask questions only when necessary. \\
\textbf{Objective:} We aim to shorten the delay before missing information is requested from the author by automatically assessing report completeness and generating context-aware questions for the missing details. \\
\textbf{Method:} We propose \tool, an iterative framework that identifies missing or ambiguous information in bug reports and generates targeted follow-up questions. Reports with no identified gaps are initially classified as sufficient. When the author responds, BRQGen evaluates whether the answers provide the requested information and generates further questions for gaps exposed by those answers, for up to five iterations. We curated a novel dataset of 429 incomplete bug reports from the Home Assistant and Brave Browser repositories, each containing at least one maintainer follow-up question. Each question is mapped to the report author’s answer and assigned a specificity label indicating whether the requested information could be provided directly through the project’s issue template.
\\
\textbf{Results:} Maintainers' early information requests were associated with the author responding, and delayed ones with no response. Among authors who responded, requests made within one hour received a median reply after 0.52 hours in Home Assistant and 1.45 hours in Brave, compared with 9.76 and 13.49 hours for those made after four hours. On the 427 incomplete reports, the generated questions covered 35.4\% (Qwen 3.8 Flash) to 51.8\% (DeepSeek V4.1 Flash) of the 1,073 follow-up questions that maintainers asked. DeepSeek V4.1 Flash managed to ask 45.7\% of all ground-truth detailed questions and 63.9\% of all generic questions. Gemini 3 Flash Preview reached the highest macro-F1 for completeness assessment of 49.3\%, but with a post-hoc analysis by getting only the gaps with high severities, Qwen 3.8 Flash reaches 73.6\% macro-F1 score. \\
\textbf{Conclusion:}
\tool~performs better at completeness assessment while at least one identified gap has high severity. The first two iterations nearly cover all ground-truth questions captured on each run compared to later iterations, including detailed questions, which cannot be directly answered from the issue template. Future work will focus on integrating context engineering via Retrieval-Augmented Generation (RAG) and validating the pipeline on industrial software ecosystems.
\end{abstract}





\maketitle

\section{Introduction}


Software maintenance is estimated to consume between 65\% and 85\% of the total resources allocated to software development \cite{ISBSG2024Maintenance}. In this phase, a bug tracking system manages the lifecycle of a defect, which typically starts with a report submitted by a tester or an end-user that the development team uses to validate, reproduce, and fix the issue \cite{torun2026}.

The time required to process a bug report varies across projects. Industrial projects typically resolve bugs within 18–27 days, whereas the average bug resolution time for the Eclipse project exceeds 300 days~\cite{zhang2017bugreport, hong2024}. An initial bottleneck is the quality of the submitted reports, which frequently lack the information required for reproducing defects.
Johnson et al. \cite{johnson2022reproduction} reported that 92\% of the reports lack all necessary S2R's to reproduce the bug.
Joorabchi et al. \cite{joorabchi2014} reported that 17\% of nearly 188k bug reports across six software repositories are non-reproducible, and that 14\% of these failures are caused directly by insufficient information. Bug reports whose descriptions contain fewer than 100 words require, on average, an additional 121 days to resolve \cite{zhang2017bugreport}. In a small-scale case study, Imran et al. \cite{imran2021} found that authors answered follow-up questions at a rate of 73\%, which suggests that follow-up questions are a practical way to obtain missing information.


Prior studies have proposed approaches that detect missing information in bug reports \cite{marcus2017, chaparro2019, bheree2024, karim2019}, but they only flag predefined elements, such as the steps to reproduce, that are missing or deficient. More recent studies use Large Language Models (LLMs) to improve bug reports \cite{bo2024, acharya2025}. However, Akyol et al. \cite{akyol2026improbr} found that executable `Steps to Reproduce' (S2R) were present in only 28\% of their real-world bug reports, and their LLM-only pipeline introduced hallucinations that raised the incidence of \textit{Wrong Information} from 2.9\% to 13.7\%. An LLM that improves a report without input from the author can therefore add incorrect information, which motivates obtaining the missing information from the author through follow-up questions \cite{suprmind2026hallucinations}.

Imran et al. \cite{imran2021} automated follow-up questions for deficient bug reports by selecting questions from a historical corpus, so their approach can ask only questions that already appear in that corpus. Generative follow-up questions have been used to retrieve missing user information in medical conversations \cite{gatto2025}, recommendation systems \cite{montazeralghaem2025}, and in software engineering \cite{walid2026}. For instance, FeedAIde \cite{walid2026} uses multimodal LLMs to generate questions during the initial issue reporting phase. However, FeedAIde doesn't adaptively assess report completeness or ask further follow-up questions only when required. To the best of our knowledge, no prior study has implemented an interactive, post-submission automated pipeline that iteratively diagnoses the semantic completeness of a submitted bug report and generates context-specific follow-up questions.


To bridge this gap, this study proposes a semi-automated framework that iteratively evaluates whether a bug report contains sufficient information and, if deficiencies are found, generates targeted, context-aware follow-up questions. This process repeats, and each new iteration evaluates the author's answers to the previous questions. We address the following research questions:

\vspace{0.1cm}

\textbf{RQ1: How is the latency of maintainers' information requests associated with author response rates and response times?} We conducted an empirical analysis on Home Assistant and Brave Browser repositories that examines whether earlier information requests from maintainers are associated with more frequent and faster author responses.

\vspace{0.1cm}

\textbf{RQ2: How accurately can the system generate follow-up questions?} We evaluate the coverage of the ground-truth questions by the generated follow-up questions and the importance ranking of the generated questions with Mean Reciprocal Rank (MRR), Normalized Discounted Cumulative Gain (NDCG), precision, and recall.

\vspace{0.1cm}

\textbf{RQ3: How effectively can the system decide whether the bug report requires additional information?} We test on both incomplete and complete bug reports how effectively the system decides whether a report is complete, so that it avoids asking the author redundant questions.

To achieve these objectives, we propose \tool, which operates as an iterative loop over a bug report. In each iteration, a single LLM call takes on two of the following three roles:
\begin{enumerate}
    \item \textbf{Bug Report Evaluator} (first iteration): analyzes the report title, body, and the comments posted before the maintainer's first follow-up question, excluding bot comments, and flags fields whose absence prevents reproduction, if any.

    \item \textbf{Answer Evaluator} (subsequent iterations): given the question--answer pairs of the previous iteration, decides for each pair whether the answer is sufficient, and derives the remaining gaps that arise from those answers, if any.

    \item \textbf{Question Generator} (every iteration): writes follow-up questions addressed to the author, only for the gaps flagged by the evaluator in the same call.
\end{enumerate}

This cycle repeats until all critical fields for reproduction are determined to be complete and unambiguous, none of the generated questions are answered, or the maximum number of iterations is reached. The main contributions of this work are:

\begin{itemize}
    
    \item \tool, a semi-automated framework that decides whether a bug report is complete to be reproduced and, if not, generates context-aware follow-up questions up to five iterations.
    
    \item A novel dataset of 429 incomplete bug reports from two repositories with 1,081 annotated maintainer questions, each labeled with its specificity (whether it can be
    answered directly from the initial bug report template) and mapped to the author's
    corresponding answer.

    \item To the best of our knowledge, this is the first evaluation framework that directly assesses the quality of generated follow-up questions for incomplete bug reports based on previously asked real-world questions by maintainers.
\end{itemize}

\section{Background and Motivation}

\begin{figure}[h]
    \centering
    \begin{subfigure}[t]{0.49\linewidth}
        \centering
        \includegraphics[width=\linewidth]{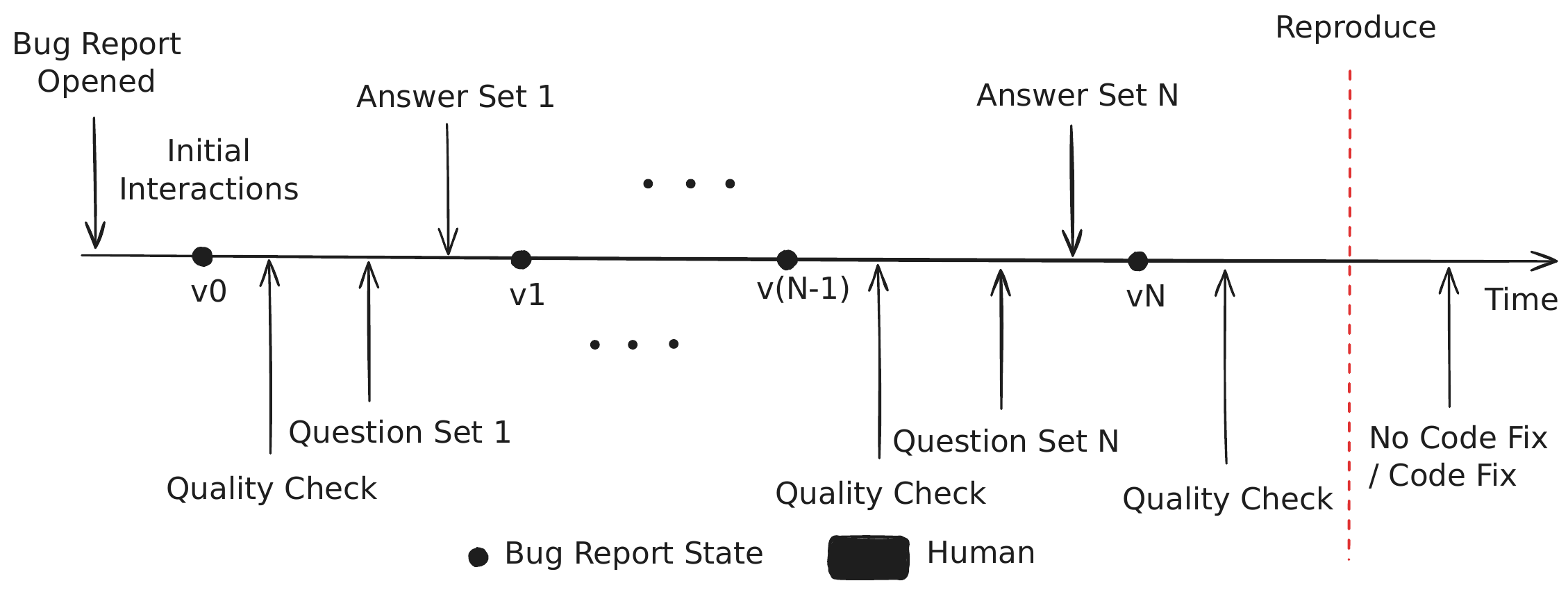}
        \caption{Typical Manual Workflow}
        \label{fig:timeline_manual}
    \end{subfigure}
    \hfill
    \begin{subfigure}[t]{0.49\linewidth}
        \centering
        \includegraphics[width=\linewidth]{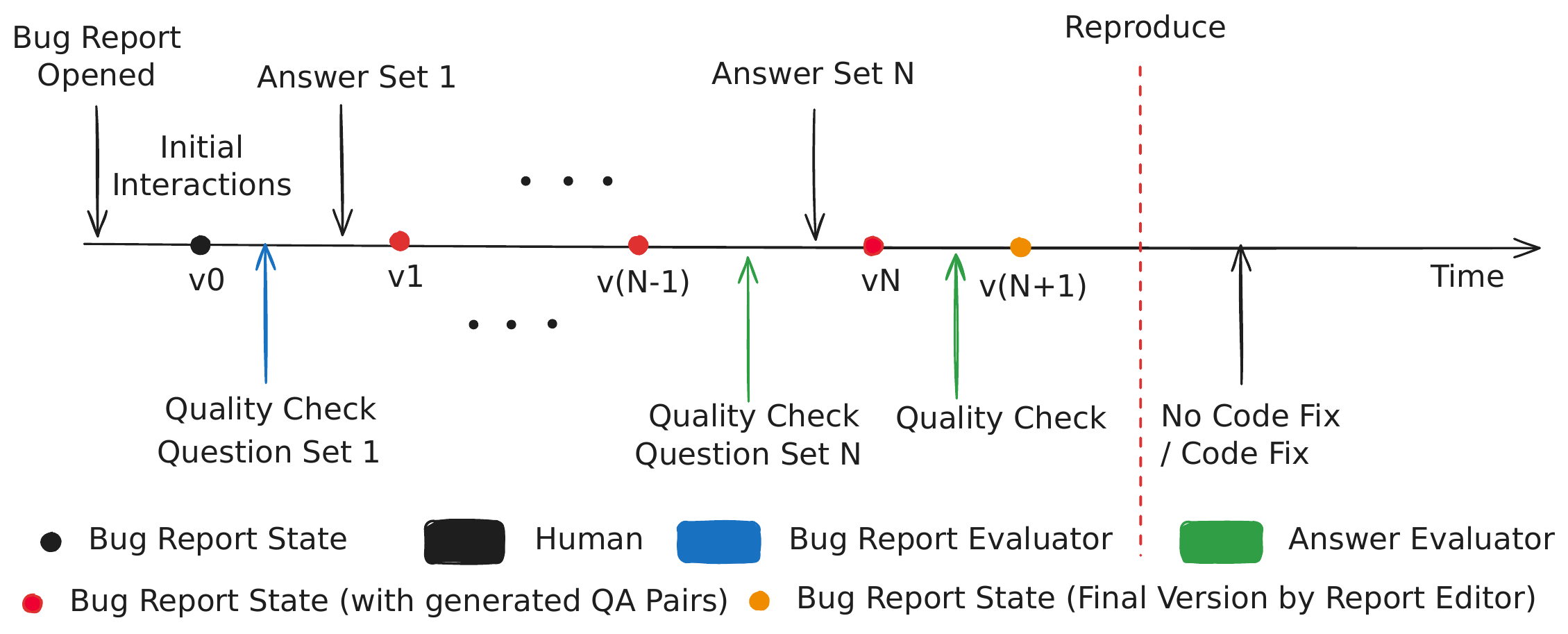}
        \caption{\tool~Workflow}
        \label{fig:timeline_agentic}
    \end{subfigure}
    \caption{Bug Report Timeline Diagrams}
    \label{fig:bug_report_timeline}
\end{figure}

The typical interaction timeline in the manual workflow begins when the author opens the bug report, in the state \textit{v0} before any maintainer-requested follow-up question. A maintainer then performs a quality check to determine whether the report is actionable and, if information is missing, asks a set of independent follow-up questions. The author's answers update the report to a new state (e.g., \textit{v1}) that includes the new question--answer pairs, and this loop continues for N iterations until the report is deemed reproducible (red dashed line). The report then proceeds to reproduction, which ends in a code fix for a valid report or in a resolution without code changes (no-code fix) for an invalid report \cite{torun2026, gon2026}. Figure \ref{fig:timeline_manual} illustrates this manual timeline. 

\begin{figure}[htpb]
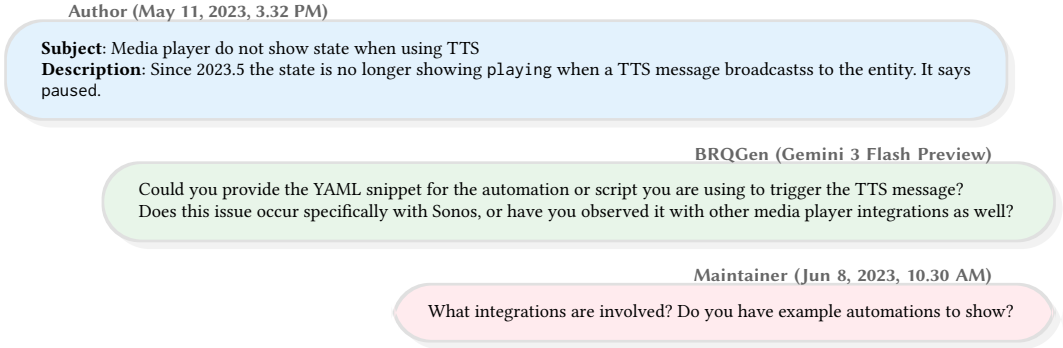

\centering


\chatLeft{Author (May 11, 2023, 3.32 PM)}{repColor}{%
\textbf{Subject}: Media player do not show state when using TTS\newline
\textbf{Description}: Since 2023.5 the state is no longer showing \texttt{playing} when a TTS message broadcastss to the entity. It says \texttt{paused}.
}
\chatRight{BRQGen (Gemini 3 Flash Preview)}{maint3Color}{%
Could you provide the YAML snippet for the automation or script you are using to trigger the TTS message?\newline
Does this issue occur specifically with Sonos, or have you observed it with other media player integrations as well?
}
\chatRight{Maintainer (Jun 8, 2023, 10.30 AM)}{maint1Color}{%
What integrations are involved? Do you have example automations to show?
}


\caption{Motivating Example of an Issue Report in Home Assistant Repository (times in UTC)}
\label{fig:chatdialogue_img}
\end{figure}

Figure~\ref{fig:chatdialogue_img} shows a motivating example from the \textit{home-assistant/core} repository\footnote{\url{https://github.com/home-assistant/core/issues/92942}}. The author states that, since Home Assistant 2023.5, a media player shows the state paused instead of playing while a text-to-speech (TTS) message is broadcast to it, but the report names only \texttt{tts} as the integration causing the issue and leaves the example YAML field empty. A maintainer asked for the integrations involved and an example automation four weeks later, the author responded one day later, and then the triage continued to match the relevant maintainer to the issue, and the cause was identified two days after the author response. Of the 32 days between the report and its closing, 28 passed before the maintainer's first information request. When run on this report, \tool~with Gemini 3 Flash Preview asked for the same information, which can be asked at submission time.

\section{Related Work}
\paragraph{Bug Report Templates}

Issue templates are sets of predefined questions and fields that guide users during issue creation and align the collected information with project expectations. Sulun et al. \cite{sulun2024} showed that projects using issue templates had significantly shorter resolution times (from 381.02 to 103.18 days) and fewer issue comments (from 4.95 to 4.32) than projects without them. Li et al. \cite{liyan2023} reported a reduction of the median resolution time from 68 to 54 days after templates were introduced, and studies by Li et al. \cite{li2023} and Zhang et al. \cite{zhang2024templates} found that template usage correlates with fewer comments. However, static templates are insufficient replacements for dynamic interaction. First, rigid templates do not adapt to the complexity of an issue and they impose unnecessary overhead on trivial reports, such as minor UI glitches, while omitting technical context required for complex issues, such as third-party service integrations. Maintainers surveyed by Sulun et al. \cite{sulun2024} noted that templates can be challenging for users who already possess the necessary information, and authors surveyed by Li et al. \cite{li2023} stated that many issues do not fit the prescribed format. Second, templates cannot resolve ambiguity or missing data through clarification. Song et al. \cite{song2022burt} found that novice authors provided on average 6.7 steps to reproduce with an interactive reporting system, compared with 3.5 with a template-based one, and authors surveyed by Li et al. \cite{li2023} named language barriers, which interactive approaches can reduce, as a challenge.

\paragraph{Automated Bug Report Quality Assessment and Improvement}

Elements such as S2R, stack traces, and test cases are frequently omitted from bug reports \cite{zimmerman2010}. Initial reports in major open-source projects rarely contain actionable data beyond observed and expected behavior \cite{davies2014}, industrial reports particularly lack application configuration and operating data \cite{laukkanen2011}, and in web applications the usage context and diagnostic details are typically absent upon submission and elicited later \cite{wang2025}. Machine learning classifiers detect the presence of Expected Behavior (EB), Observed Behavior (OB), S2R, and valid stack traces across various languages and datasets \cite{marcus2017, karim2019, song2020, bheree2024}. LLM-based approaches go further and generate the missing report components; as ChatBR \cite{bo2024} combines a fine-tuned BERT classifier with ChatGPT to iteratively synthesize absent content, ImproBR \cite{akyol2026improbr} uses a Retrieval-Augmented Generation (RAG) pipeline to produce context-aware enhancements, Acharya and Ginde \cite{acharya2025} show that instruction fine-tuned models can transform unstructured descriptions into well-formatted reports, and, for graphical user interfaces (GUIs), AstroBR \cite{mahmud2025} and BugScribe \cite{saha2026automated} combine LLMs with app-specific execution models to evaluate semantic accuracy and synthesize missing interface interactions. All of these approaches generate improvements without feedback from the author, which risks hallucination when the initial report lacks the necessary context.

\paragraph{Interactive Bug Report Improvement}

Imran et al.~\cite{imran2021} proposed selecting follow-up questions from a historical corpus, which limits the questions to those asked about earlier reports. FUSION~\cite{moran2015fusion}, EBug~\cite{9773038}, and BURT~\cite{song2022burt} provide real-time assistance as users write their reports, but only for specific structural elements such as S2R or Android GUI interactions. FeedAIde~\cite{walid2026} uses multimodal LLMs to capture contextual logs and generate adaptive questions during the reporting phase, and recent industrial workflows~\cite{leung2026} embed AI assistants at the point of discovery that guide the author through preliminary triage before submission. These approaches guide the author while the report is being written. In contrast, our framework acts as a post-submission automated maintainer that assesses report completeness and queries the author only when critical information is missing.

\paragraph{Follow-up Question Generation in Other Domains}

Outside software maintenance, follow-up questions are used to resolve ambiguity and extract missing context. The medical dialogue systems generate personalized questions to clarify patient-reported conditions \cite{gatto2025}, recommendation frameworks ask sequential clarifying questions to reconstruct incomplete user profiles \cite{montazeralghaem2025}, and conversational search engines ask questions to disambiguate short, unstructured queries \cite{zamani2020, aliannejadi-etal-2021-building}; for software developers, clarification questions elicit additional terms for unsuccessful Web search queries \cite{imran2022clarification}. In code generation, ClarifyGPT asks clarifying questions only when code sampled for the same requirement behaves inconsistently \cite{mu2024clarifygpt}, a signal that a bug report cannot provide because it cannot be executed. In requirements engineering, dynamic question generation elicits tacit user needs \cite{shen2025}, and Wood et al. \cite{wood2018} showed that human-to-human bug repair conversations rely on clarifying dialogue.

\section{Methodology}
\begin{figure*}
    \centering
    \includegraphics[width=1\linewidth]{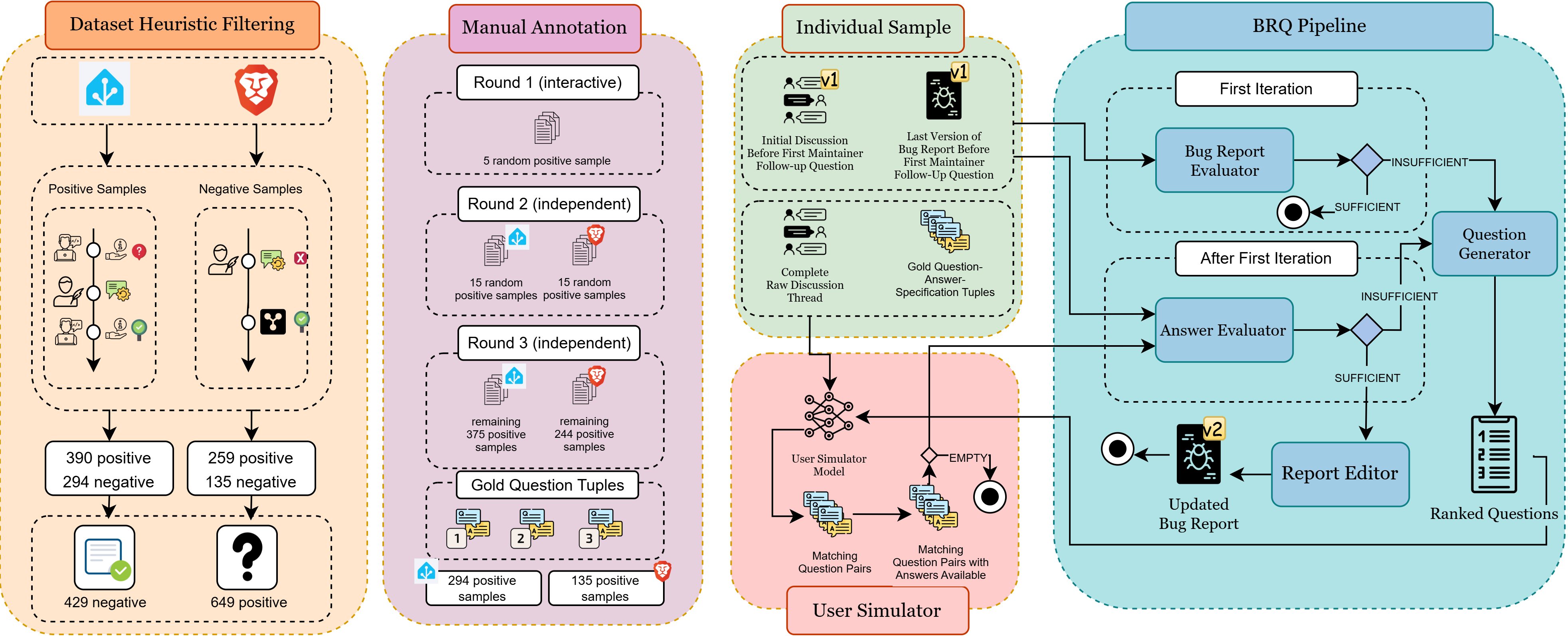}
    \caption{High Level Diagram for \tool}
    \label{fig:brqgen}
\end{figure*}

In this section, we present our dataset preparation strategy, the high-level architecture of the system, and how this architecture changes the bug report timeline shown in Figure~\ref{fig:timeline_manual}. \Cref{fig:brqgen} summarizes the overall methodology.

\subsection{Dataset Collection and Preprocessing}
\label{sec:dataset}

\subsubsection{Repository Selection}
\label{sec:repo_selection}

To evaluate \tool~beyond a single project, we selected two repositories based on the following criteria: (i) a large user base with diverse technical skill levels, (ii) active user-maintainer discussions, (iii) a sufficient number of closed issues to ensure a reliable ground truth dataset, and (iv) the presence of an issue label implying that the maintainers require additional information from the author (\textit{more-information}). The first repository, \textit{home-assistant/core}\footnote{https://github.com/home-assistant/core}, has installation paths that span every level of technical skill, from beginners to advanced users requiring command-line proficiency\footnote{\url{https://www.home-assistant.io/installation/}}. As of September 20th, 2026, it had 90.9k stars and 71,527 closed issue reports, 2,742 of which received its \textit{more-information} label, \textit{needs-more-information}. The second repository, \textit{brave/brave-browser}\footnote{https://github.com/brave/brave-browser}, offers easy install options for beginners\footnote{\url{https://brave.com/download/}} and customizable client-side options from its open-source resources for advanced users. It had 23.7k stars and 45,069 closed issue reports, 809 of which received its \textit{more-information} label, \textit{needs-more-info}.

\subsubsection{Candidate Dataset Construction}

After selecting the repositories, we constructed two candidate datasets. Candidate positive samples (\positiveDataset) are reports likely to contain a maintainer follow-up question and a subsequent author response, whereas negative samples (\negativeDataset) are reports likely to have required no further clarification. Both are selected with the heuristics described below, and the candidate positive samples are then validated through manual annotation.


\paragraph{Positive-set heuristic.}


To construct the candidate positive set, we applied strict heuristic filtering to the closed issues of both repositories. We required that the \textit{more-information} label was applied by a maintainer and subsequently removed, and that the issue author commented while the label was active. This signal increases the likelihood that the label represented a genuine information gap and that the author attempted to provide the requested information. The filtering produced 259 candidate reports from \textit{brave/brave-browser} and 390 from \textit{home-assistant/core}, for a total of 649. This candidate set is relatively small because the \textit{more-information} label is rare in both repositories (see Section~\ref{sec:repo_selection}); we retained these strict constraints to reduce noise before manual annotation.

\paragraph{Negative-set heuristic.}

For the \negativeDataset~split, we retained a closed bug report only if (i) the author never commented on it, (ii) the author never edited the bug report after its opening, (iii) it never received the \textit{more-information} label, and (iv) it was closed through a merged pull request or a commit. We scanned the closed issues of each repository from the most recently created one and stopped when the number of retained reports matched the size of the final annotated positive set in that repository (see Section~\ref{sec:final_dataset_statistics}), which yielded 429 reports, with 294 from \textit{home-assistant/core} and 135 from \textit{brave/brave-browser}.

\subsubsection{Annotation Schema}
\label{sec:annotation_schema}

Initially, the ground-truth follow-up questions of the candidate \positiveDataset\ samples and their answers are not explicitly mapped. We therefore manually annotate each candidate report to construct a set of question--answer mappings $M_{QA} = \{(q_k, A_k, s_k)\}_{k=1}^{K}$, where $K$ is the total number of ground-truth follow-up questions identified in the report and each tuple represents one annotated follow-up question:

\begin{itemize}
\item $q_k$ is the textual copy of the $k$-th ground-truth follow-up question that a maintainer asked the issue author, as it appears in the original comment thread.

\item $A_k = \{a_{k,1}, \ldots, a_{k,n}\}$ is the set of textual copies of the issue author's comments that answer $q_k$; $A_k = \emptyset$ when no answer to $q_k$ is provided, and $|A_k| > 1$ when the answer is distributed across multiple comments.

\item $s_k \in \{\textsc{Generic}, \textsc{Detailed}\}$ is the specificity of $q_k$: \textsc{Generic} if $q_k$ can be directly addressed using information requested by the project-specific issue template, and \textsc{Detailed} otherwise.
\end{itemize}

\subsubsection{Annotation Protocol}

Three authors annotated the 649 candidate positive reports. To establish an initial annotation guideline, they first jointly annotated five randomly selected reports. After drafting the guideline, all three annotators independently annotated a calibration set of 30 reports (15 from each repository) that included these five, and they discussed the disagreements across the 30 independent annotations to finalize the guideline. The final guideline consists of 13 rules for question extraction, answer extraction, and specificity labeling, and it is available in our replication package (\Cref{sec:data_availability}). The remaining 619 candidate reports were partitioned into three disjoint batches, and each annotator was assigned two of them, so that every report received exactly two independent annotations. When the paired annotations disagreed, all three annotators jointly reviewed the report and resolved the disagreement through majority voting.

\subsubsection{Inter-rater Agreement}

Inter-rater agreement was measured separately for question extraction, answer extraction, and specificity classification. Because annotators may identify different numbers of questions for the same report and may order them differently, we measured question-text overlap using the \textit{Dice--Sørensen coefficient (DSC)} with exact string matching. The resulting question-text agreement was .6349 for micro DSC and .7069 for macro DSC; the higher macro score reflects reports for which both annotators agreed that no eligible follow-up question was present. After matching questions between annotators, we obtained 600 matched question pairs, on which the answer-text agreement measured by the \textit{exact matching ratio} was 69.83\% and the specificity agreement reached a Cohen's kappa of .6432, indicating substantial agreement.

\subsubsection{Final Dataset Statistics}
\label{sec:final_dataset_statistics}

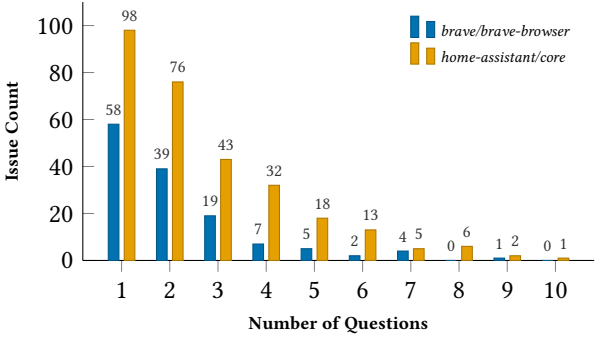
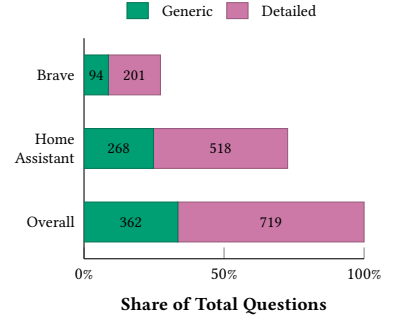
\begin{figure}[t]
    \centering
    \begin{subfigure}[c]{0.60\textwidth}
        \centering
        \begin{tikzpicture}
            \begin{axis}[
                ybar,
                width=\linewidth,
                height=5.0cm,
                ymin=0,
                ymax=110,
                xmin=0.4,
                xmax=10.6,
                ylabel={\textbf{Issue Count}},
                xlabel={\textbf{Number of Questions}},
                x label style={font=\scriptsize},
                y label style={font=\scriptsize},
                xtick={1,2,3,4,5,6,7,8,9,10},
                ytick={0,20,40,60,80,100},
                axis lines*=left,
                bar width=4pt,
                enlarge x limits=0.02,
                nodes near coords,
                every node near coord/.append style={
                    font=\tiny,
                    color=black!80,
                    /pgf/number format/fixed
                },
                legend style={
                    at={(0.98,0.97)},
                    anchor=north east,
                    draw=none,
                    fill=white,
                    font=\tiny,
                    row sep=-1pt
                },
                legend cell align={left}
            ]

            \addplot[fill=repoBrave, draw=repoBrave!75!black] coordinates {
                (1,58) (2,39) (3,19) (4,7) (5,5)
                (6,2)  (7,4)  (8,0)  (9,1) (10,0)
            };
            \addlegendentry{\textit{brave/brave-browser}}

            \addplot[fill=repoHomeAssistant, draw=repoHomeAssistant!75!black] coordinates {
                (1,98) (2,76) (3,43) (4,32) (5,18)
                (6,13) (7,5)  (8,6)  (9,2)  (10,1)
            };
            \addlegendentry{\textit{home-assistant/core}}

            \end{axis}
        \end{tikzpicture}
        \caption{Report distribution by number of questions.}
        \label{fig:qa_details_by_repo}
    \end{subfigure}%
    \hfill
    \begin{subfigure}[c]{0.38\textwidth}
        \centering
        \begin{tikzpicture}
            \begin{axis}[
                xbar stacked,
                width=\linewidth,
                height=4.5cm,
                xmin=0,
                xmax=100,
                ymin=0.5,
                ymax=3.5,
                bar width=15pt,
                ytick={1,2,3},
                yticklabels={Overall,Home\\Assistant,Brave},
                yticklabel style={font=\tiny, align=right},
                xtick={0,50,100},
                xticklabel={\pgfmathprintnumber{\tick}\%},
                xticklabel style={font=\tiny},
                xlabel={\textbf{Share of Total Questions}},
                x label style={font=\scriptsize},
                axis lines*=left,
                point meta=explicit symbolic,
                nodes near coords,
                every node near coord/.append style={
                    font=\tiny,
                    anchor=center,
                    color=black
                },
                legend style={
                    at={(0.5,1.03)},
                    anchor=south,
                    draw=none,
                    fill=white,
                    font=\tiny,
                    legend columns=-1,
                    column sep=3pt
                },
                legend cell align={left}
            ]

            \addplot[fill=specificityGeneric, draw=specificityGeneric!75!black] coordinates {
                (33.5,1) [362] (24.8,2) [268] (8.70,3) [94]
            };
            \addlegendentry{Generic}

            \addplot[fill=specificityDetailed, draw=specificityDetailed!75!black] coordinates {
                (66.5,1) [719] (47.9,2) [518] (18.6,3) [201]
            };
            \addlegendentry{Detailed}
            \end{axis}
        \end{tikzpicture}
        \caption{Repository contributions to the overall specificity counts.}
        \label{fig:specificity_totals}
    \end{subfigure}

    \caption{Final positive-dataset distributions. Panel (a) reports the number of reports containing one through ten eligible follow-up questions. Panel (b) reports generic and detailed question counts as percentages of all 1,081 annotated questions; values inside the bars are absolute question counts.}
    \label{fig:qa_details}
\end{figure}

Of the 649 candidate positive reports, 220 contained no eligible follow-up question under the finalized annotation guidelines and were excluded. The final positive dataset therefore contains 429 reports, 294 from \textit{home-assistant/core} and 135 from \textit{brave/brave-browser}, with 1,081 annotated questions, 786 and 295, respectively. Of these questions, 362 (33.5\%) are generic and 719 (66.5\%) are detailed, and 870 (80.48\%) included answers (657 of 786 (83.59\%) in \textit{home-assistant/core} and 213 of 295 (72.20\%) in \textit{brave/brave-browser}). \Cref{fig:qa_details} summarizes the report-level question-count distribution and each repository's contribution to the generic and detailed question counts.

\subsection{System Architecture}
We propose \tool, a semi-automated pipeline in which a single LLM call per iteration takes on two roles: the \textit{Bug Report Evaluator} at the first iteration, or the \textit{Answer Evaluator} at later iterations if the author answered at least one question of the previous iteration, each followed by the \textit{Question Generator}, which writes follow-up questions only for the gaps that the evaluator lists. For evaluation, a \textit{User Simulator} acts as the actual author and matches each generated question to the ground-truth maintainer questions that request the same information and returns the author's historical answers to the matched questions, which the \textit{Answer Evaluator} checks in the next iteration. As a future vision, a Report Editor module may execute an automated API call to the host issue tracking system (e.g., GitHub) to append the newly collected QA pairs to the original bug report. Figures~\ref{fig:brqgen} and \ref{fig:timeline_agentic} detail the system's activity and timeline. 



\subsubsection{Bug Report Evaluator}

The \textit{Bug Report Evaluator} determines whether a bug report, in its state before the first follow-up question, contains sufficient information for the bug to be reproduced, and formulates the first iteration of follow-up questions when it does not. Within a single LLM invocation, its underlying evaluator model $\mathcal{M}^{\text{br}}_{\text{eval}}$ enumerates the information gaps in the report and then formulates follow-up questions restricted to those gaps.

The report is evaluated in its latest state before the first annotated follow-up question $q_1$ was asked, denoted $B_0 = (b, t, C_0, u, D_0)$, where $b$ is the report body at that time, $t$ its title, $u$ its author, $C_0$ the comments posted before the comment that contains $q_1$, and $D_0$ the additional media referenced in $b$ and $C_0$. We remove bot comments from $C_0$ and mark each remaining comment as written by the author, maintainer or another participant, so that the model can tell the author's additions from maintainers and other people's comments. Negative reports have no follow-up question, so their $C_0$ is empty. The media in $D_0$, such as screenshots, logs, or linked files, are resolved and supplied to the model alongside the textual input.

The model utilizes $B_0$ and returns an evaluation tuple $\omega_1$ which is the set of identified missing or
ambiguous fields, together with an ordered question list $Q_1$ based on severity:

\begin{equation}
    \bigl(\omega_1, Q_1\bigr)
    = \mathcal{M}^{\text{br}}_{\text{eval}}(B_0),
    \qquad
    \omega_1 =
    \{(f_k, r_k, \sigma_k)\}_{k=1}^{N}.
\end{equation}

where $f_k$ is the field name, $r_k$ the justification for its
absence, and $\sigma_k \in \{\text{LOW}, \text{MEDIUM}, \text{HIGH}\}$ the
severity of the omission. The generated question list is $Q_1 = [q_1^{(1)}, \dots, q_1^{(N_q)}]$, where each $q_1^{(m)}$ is instructed to target exactly one field $f_k$ in $\omega_1$. Notably, the model is not asked to directly classify the report.
Instead, the binary completeness label is derived deterministically from the
identified gap set:

\begin{equation}
    c_1 =
    \begin{cases}
        \text{SUFFICIENT}   & \text{if } \omega_1 = \emptyset, \\
        \text{INSUFFICIENT} & \text{otherwise}.
    \end{cases}
\end{equation}

This decoupling prevents self-inconsistent outputs in which the model declares a report sufficient while simultaneously identifying missing information.

We instruct $\mathcal{M}^{\text{br}}_{\text{eval}}$ to adhere to three constraints through explicit prompt instructions:

\begin{itemize}
    \item The model is explicitly prohibited from flagging information
    whose absence could only be resolved by asking the author to perform environment modifications that require source code changes to try to directly resolve the issue
    (e.g., \textit{Can you update to the latest version and see if it works?}).

    \item Gaps within $\omega_1$ must be mutually independent. The model may not request information whose relevance depends on the answer to another gap raised at the same iteration.

    \item Questions may be generated only for gaps enumerated in $\omega_1$, so that every element of $Q_1$ is traceable to an explicitly justified omission, and each question should only address one identified gap.
\end{itemize}

To prioritize the most important missing information, $Q_1$ is ranked in descending order of severity, with reproducibility-critical gaps prioritized within equal-severity groups. $N_q$ is not fixed, since the model initially emits one question for each identified gap, after which the list is truncated to the $N_{\max} = 10$ highest-ranked questions to reach the upper bound ground-truth question count of one sample in the positive dataset (see Fig.~\ref{fig:qa_details_by_repo}).

We pose the selected questions together rather than one at a time. This is
possible because gaps raised at the same iteration are required to be
independent; for example, the operating system version can be requested
independently of a stack trace. Asking these questions together is intended to
reduce the number of interaction rounds and the total author response latency.

\subsubsection{Answer Evaluator}

The \textit{Answer Evaluator}, with model $\mathcal{M}^{\text{ans}}_{\text{eval}}$, handles iterations $\ell \in \{2, \dots, L_{\max}\}$, where $L_{\max}$ is the maximum number of iterations, and runs only when at least one question of iteration $\ell-1$ received an answer. It can request the information that the \textit{Bug Report Evaluator} may not request, which is the information whose relevance depends on the answer to another gap. It therefore does not re-evaluate the report; it checks whether each received answer provides the requested information and lists the gaps these answers expose. Let
\begin{equation}
    F_{\ell-1}
    =
    [(q^{(1)}, a^{(1)}), \dots, (q^{(n)}, a^{(n)})]
\end{equation}
denote the \textit{frontier} at iteration $\ell$: the $n$ pairs of iteration $\ell-1$ whose questions were answered by the author. As in breadth-first search, the frontier holds the most recently answered pairs from which the next gaps are derived; unanswered questions are not carried forward. The model receives the frontier and, as background, the same $B_0$, but not the pairs of earlier iterations, since we assume that the report will be complete if all of the generated questions are answered sufficiently, therefore we reduce the problem from completeness of the bug report to the completeness of the answers of the questions:
\begin{equation}
    \bigl(\alpha_\ell, G_\ell, Q_\ell\bigr)
    =
    \mathcal{M}^{\text{ans}}_{\text{eval}}(B_0, F_{\ell-1}).
\end{equation}

$\alpha_\ell = \{(i, \text{suf}_i)\}_{i=1}^{n}$ records whether answer $a^{(i)}$ provides the information requested by $q^{(i)}$. $G_\ell = \{(f_k, r_k, \sigma_k, P_k)\}_{k=1}^{N}$ contains the \textit{related gaps} that the \textit{Answer Evaluator} identifies; each has the fields of a gap in $\omega_1$ plus a non-empty parent set $P_k \subseteq \{1, \dots, n\}$ of the frontier pairs it arises from. An insufficient answer leads to a related gap on its pair unless nothing more can be asked about it, and an answer can also make new information relevant. $Q_\ell$ contains questions only for these gaps, and each question carries its gap's parent set, which traces it to the answers it builds on. The constraints, severity ranking, and $N_{\max}$ limit of the \textit{Bug Report Evaluator} also apply. As in the first iteration, completeness of bug report is derived using Eq. \ref{eq:cl}
\begin{equation}
\label{eq:cl}
    c_\ell =
    \begin{cases}
        \text{SUFFICIENT}
        & \text{if } G_\ell = \emptyset
          \text{ and } \text{suf}_i \text{ for all } i, \\
        \text{INSUFFICIENT}
        & \text{otherwise}.
    \end{cases}
\end{equation}

\tool~stops at iteration $\ell$ when any of the following
conditions is met: the current state is classified as sufficient; none of the generated questions receives an answer
($F_\ell = \emptyset$); or the maximum number of iterations $L_{\max}$ is reached.
In our experiments, we set $L_{\max}=5$ and $N_{\max}=10$.

\subsubsection{User Simulator}

Following Montazeralghaem et al.~\cite{montazeralghaem2025} and Mu et al.~\cite{mu2024clarifygpt}, we implement a fully automated \textit{User Simulator} ($\mathcal{M}_{\mathrm{sim}}$) to replace the costly process of obtaining  multi-turn responses from authors. While \tool~ has access only to $B_0$ and to the answers returned by the simulator, $\mathcal{M}_{\mathrm{sim}}$ acts as an oracle over the ground-truth $M_{QA}$ mappings defined in Section~\ref{sec:annotation_schema}.

At iteration $\ell$, the pipeline generates a set of follow-up questions $Q_\ell$. The complete set is provided to the simulator in a single call, which identifies, for each generated question, the indices of ground-truth questions that directly and specifically request the same information. Importantly, only the ground-truth questions $Q_{GT}=\{q_1,\ldots,q_K\}$ are exposed during this matching stage; their corresponding answers are withheld until the matching decisions are finalized. This prevents tangential information contained in historical answers from producing false-positive matches.

The simulator is instructed to reject matches based solely on broad topical similarity, shared terminology, or partial relevance. To reduce sampling variance, we execute the matching procedure three times on the same $Q_\ell$ and retain, for each generated question, only the ground-truth indices selected in at least two of the three runs. The available answers of the retained ground-truth questions are then retrieved deterministically and concatenated in ascending index order; if none of them has an available answer, including the case where no ground-truth question matches, the simulator returns \texttt{NO\_RESPONSE}. The simulator also returns the retained indices, so that a matched question without an available answer remains distinguishable from a question with no ground-truth match. Once a generated question has been matched to a ground-truth question and answered, the simulator no longer offers that ground-truth question for matching in later iterations.

We instantiate $\mathcal{M}_{\mathrm{sim}}$ using \texttt{openai/gpt-5.6-terra} \cite{openai2026gpt56} with medium reasoning effort and the default sampling temperature, with the same configuration and matching prompt in all three runs.

\paragraph{Manual Alignment Evaluation}
\label{sec:user-simulator-alignment}
We measured how well User Simulator agrees with human judgment. We ran \tool~on a sample of positive reports with the authors in the role of the User Simulator, and then gave the same generated questions to the User Simulator.

\textit{Sampling.} We sized the sample in ground-truth questions but drew whole reports, since questions are generated only by running \tool~on a report and the User Simulator matches all of a report's questions in one call. Cochran's formula with finite population correction~\cite{cochran1977} (95\% confidence, 10\% margin of error, $p = 0.5$) gives 88.28 of the 1,081 ground-truth questions. To account for questions within a report being correlated, we multiplied this by Kish's design effect~\cite{kish1965} of 1.59, based on a size-weighted mean of 3.78 questions per report and an intra-class correlation of 0.214, which we estimated from the questions' wording because matching outcomes were unknown before annotation, yielding a target of 141 questions. We stratified by repository in proportion to its ground-truth questions (38 Brave, 103 Home Assistant) and, within each, added reports in random order (seed 42) with all their questions until the quota was met. The sample contains 59 reports with 144 questions: 19 Brave reports (39 questions) and 40 Home Assistant reports (105 questions).

\textit{Human labels and scoring.} Each sampled report was run with one of the three models, assigned at random and balanced within each repository by the number of ground-truth questions: Qwen 3.8 Flash~\cite{alibaba2026qwen38flash} and DeepSeek V4.1 Flash~\cite{deepseek2026v41flash} on 20 reports each and Gemini 3 Flash Preview~\cite{google2025gemini3flash} on 19. In these runs, the three authors replaced the User Simulator and for each generated question, they jointly selected, until agreement, the ground-truth questions that directly and specifically request the same information, the same criterion given to the User Simulator. We call each selected pair of a generated and a ground-truth question a \textit{match}; a generated question can have any number of matches. Qwen 3.8 Flash judged one Home Assistant report complete and asked no question, leaving 58 reports with 143 ground-truth questions, on which the runs produced 259 generated questions and the authors made 55 matches. We then gave the same generated questions to the User Simulator, one iteration at a time, offering only the ground-truth questions still open in the authors' run so that disagreements do not propagate across iterations.

\textit{Results.} The User Simulator made 57 matches, 47 of which the authors also made: it missed 8 of the authors' 55 matches and added 10 that the authors did not make. This gives a precision of 0.82 (share of the User Simulator's matches that the authors also made), a recall of 0.85 (the share of the authors' matches that the User Simulator also made), and an $F_1$ score of 0.84. The $F_1$ score was similar for all three models, from 0.83 to 0.84. Because the missed and the added matches nearly balance, the coverage reported for RQ2 is close to the coverage under human judgment.



\subsection{Experimental Setup}
\label{sec:experimental_setup}

To evaluate \tool, we used three LLMs through OpenRouter: Gemini 3 Flash Preview, Qwen 3.8 Flash, and DeepSeek V4.1 Flash. As of September 2026, these were the most recent models that accept a temperature of zero with reasoning disabled. We ran every \tool~call with temperature zero and reasoning disabled, to bring decoding as close to deterministic as the model providers allow. Two positive home-assistant/core reports ended with parse error on DeepSeek V4.1 Flash. We exclude these two reports for all three models, so that we evaluate every model on the same 427 positive and 429 negative reports in RQ2 and RQ3. The User Simulator runs with medium reasoning enabled and at the model's default sampling temperature, as described in Section~\ref{sec:user-simulator-alignment}.

\textbf{\textit{RQ1: Maintainers' Information Request Latency and Author Responses.}} To examine how the timing of a maintainer's information request is associated with author engagement, we conducted two empirical analyses on home-assistant/core and brave/brave-browser. The first analysis investigated the relationship between the latency of a maintainer's information request and the author's likelihood to reply. We gathered the issues where the \textit{more-information} label was assigned by a non-author, 2,742 for Home Assistant and 809 for Brave, and categorized them into three cohorts: (1) the author never responded, (2) the author responded but the label remained, and (3) the author responded and the label was subsequently removed. We then compared the average time elapsed before label assignment across the cohorts to determine whether delayed information requests are associated with author abandonment.

The second analysis assessed whether maintainers' earlier information requests are associated with faster author replies. We selected the closed bug reports in which the author replied after the \textit{more-information} label was assigned, 1,883 for Home Assistant and 537 for Brave, and extracted two temporal metrics for each: the Labeling Time Elapse (duration from issue creation to label assignment) and the Author Response Time Elapse (duration from label assignment to the author's first subsequent reply). The Mann--Whitney U tests and Cliff's $\delta$ are rank-based, so we computed them on all issues of both analyses. We excluded outliers in the Labeling Time Elapse with the Tukey method~\cite{tukey1977exploratory} only from the average times of the first analysis and from the trend of the second analysis, 
leaving 1,560 and 443 samples for the trend. To show the trend in the noisy response data of the second analysis, we applied Locally Weighted Scatterplot Smoothing (LOWESS) \cite{cleveland1979} 
and estimated a 95\% confidence band around each trend from 1,000 bootstrap resamples.

\textbf{\textit{RQ2: Question Generator Performance.}} We evaluated the Question Generator on the 427 positive reports, because only these reports have ground-truth follow-up questions. They contain 1,073 ground-truth questions, of which 360 are generic and 713 are detailed. A ground-truth question is \textit{covered} when the User Simulator matches at least one generated question to it in any iteration. The coverage of a report is the share of its ground-truth questions that are covered. A report is \textbf{fully} covered when all of its ground-truth questions are covered, \textbf{not} covered when none of them is, and \textbf{partially} covered otherwise. We summarize coverage as the \textit{pooled coverage}, the number of covered ground-truth questions of all reports divided by the number of their ground-truth questions, so that every ground-truth question counts once. Because reports with more ground-truth questions may be harder to cover completely, we also group the reports by their number of ground-truth questions and compute the pooled coverage and the fully, partially, and not covered reports of each group. We further attribute each covered ground-truth question to the iteration in which it was first matched, which shows how much coverage the iterations after the first add. For each iteration, we also count the ground-truth questions that were still available to the User Simulator at its start, and we compute coverage separately for generic and detailed ground-truth questions (Section~\ref{sec:annotation_schema}).

To evaluate how the Question Generator ranks its questions, we take the questions it generated for a report in one iteration, at most $N_{\max}=10$, in the order in which it ranked them by the severity of their gaps, and evaluate the first $k \in \{1, 3, 5\}$ of them. A generated question is \textit{correct} when the User Simulator matched it to at least one ground-truth question. Precision@$k$ (P) is the share of correct questions among the first $k$; when fewer than $k$ questions were generated, the missing ones count as incorrect. Recall@$k$ (R) is the number of distinct ground-truth questions matched by the first $k$ questions, divided by all ground-truth questions of the report, including those that an earlier iteration already matched. NDCG@$k$ uses a binary gain and the discount $1/\log_2(\mathit{rank}+1)$, and credits a question only when it matches a ground-truth question that no higher-ranked question matched, so that NDCG cannot exceed 1. MRR@$k$ is the reciprocal rank of the first correct question among the first $k$, or 0 when none is correct. For each iteration, we average the four metrics over the reports that reached the iteration and on which \tool~generated at least one question in it. We report iterations 1 and 2 only, because at most 28 reports per model reached iteration 3.

\textbf{\textit{RQ3: Bug Report Evaluator Performance.}} We evaluated whether the Bug Report Evaluator of \tool~correctly decides that an initial bug report needs additional information. For each of the 427 positive and 429 negative reports, we took the completeness verdict that the Bug Report Evaluator produced at the first iteration. The Bug Report Evaluator labels a report insufficient when it lists at least one missing or ambiguous field and sufficient when it lists none. The correct verdict for a positive report is insufficient, and the correct verdict for a negative report is sufficient. We report the share of positive reports judged insufficient, the share of negative reports judged sufficient, and the macro-F1 to demonstrate the overall performance. In a supplementary post-hoc analysis, we also derive the verdict from the severities of the listed fields ($\sigma_k$) with a severity threshold: a report is insufficient when at least one of its fields has the threshold severity or a higher one, and sufficient otherwise. The threshold \textit{low} is the verdict of \tool, \textit{medium} judges a report whose fields all have the severity low as sufficient, and \textit{high} considers a report insufficient, only if it contains at least one field with high severity. Because a report judged sufficient receives no questions, we also report the RQ2 coverage that each threshold retains (Section~\ref{sec:rq3_results}).
\subsection{Generative AI Usage}

We used Claude Code (Claude Opus~5 \cite{anthropic2026opus5} and~5.5 \cite{anthropic2026opus55}) as a coding assistant for
parts of the evaluation pipeline, the sampling script and the analysis
notebooks. The authors reviewed and tested all generated code and checked
every reported number against the notebooks' output.

\section{Results}

\subsection{Empirical Study}

\begin{figure}[t!]
    \centering

\begin{subfigure}[b]{0.48\linewidth}
    \centering
    \begin{tikzpicture}
        \begin{axis}[
            xbar,
            width=\linewidth,
            height=3.5cm,
            xmin=0, xmax=190,
            ymin=0.5, ymax=3.5,
            xtick={0,50,100,150},
            ytick={1,2,3},
            yticklabels={
                {Author Did\\Not Respond},
                {Author Responded but\\The Label Stayed},
                {Author Responded and\\The Label Removed}
            },
            yticklabel style={font=\tiny,align=right},
            xticklabel style={font=\tiny},
            xlabel={Elapsed Time of Maintainer Information Request},
            xlabel style={font=\scriptsize},
            axis lines*=left,
            bar width=11pt,
            bar shift=0pt
        ]
            \addplot[
                fill=blue!65,
                nodes near coords={7.0h (448)},
                every node near coord/.append style={
                    font=\tiny,anchor=west,xshift=2pt
                }
            ] coordinates {(7.0,3)};

            \addplot[
                fill=orange!75,
                nodes near coords={9.8h (1116)},
                every node near coord/.append style={
                    font=\tiny,anchor=west,xshift=2pt
                }
            ] coordinates {(9.8,2)};

            \addplot[
                fill=green!60!black,
                nodes near coords={160.6h (738)},
                every node near coord/.append style={
                    font=\tiny,anchor=east,xshift=-2pt,
                    fill=white,inner sep=1pt
                }
            ] coordinates {(160.6,1)};
        \end{axis}
    \end{tikzpicture}
    \caption{HA: Average time per semantic cluster.}
    \label{fig:ha_time_plot}
\end{subfigure}
    \hfill
\begin{subfigure}[b]{0.48\linewidth}
    \centering
    \begin{tikzpicture}
        \begin{axis}[
            xbar,
            width=\linewidth,
            height=3.5cm,
            xmin=0, xmax=650,
            ymin=0.5, ymax=3.5,
            xtick={0,200,400,600},
            ytick={1,2,3},
            yticklabels={
                {Author Did\\Not Respond},
                {Author Responded but\\The Label Stayed},
                {Author Responded and\\The Label Removed}
            },
            yticklabel style={font=\tiny,align=right},
            xticklabel style={font=\tiny},
            xlabel={Elapsed Time of Maintainer Information Request},
            xlabel style={font=\scriptsize},
            axis lines*=left,
            bar width=11pt,
            bar shift=0pt
        ]
            \addplot[
                fill=blue!65,
                nodes near coords={193.4h (247)},
                every node near coord/.append style={
                    font=\tiny,anchor=west,xshift=2pt
                }
            ] coordinates {(193.4,3)};

            \addplot[
                fill=orange!75,
                nodes near coords={112.1h (212)},
                every node near coord/.append style={
                    font=\tiny,anchor=west,xshift=2pt
                }
            ] coordinates {(112.1,2)};

            \addplot[
                fill=green!60!black,
                nodes near coords={555.9h (244)},
                every node near coord/.append style={
                    font=\tiny,anchor=east,xshift=-2pt,
                    fill=white,inner sep=1pt
                }
            ] coordinates {(555.9,1)};
        \end{axis}
    \end{tikzpicture}
    \caption{Brave: Average time per semantic cluster.}
    \label{fig:brave_time_plot}
\end{subfigure}

    \vspace{0.5em}

    \begin{subfigure}[b]{0.48\linewidth}
        \centering
        \includegraphics[width=\linewidth]{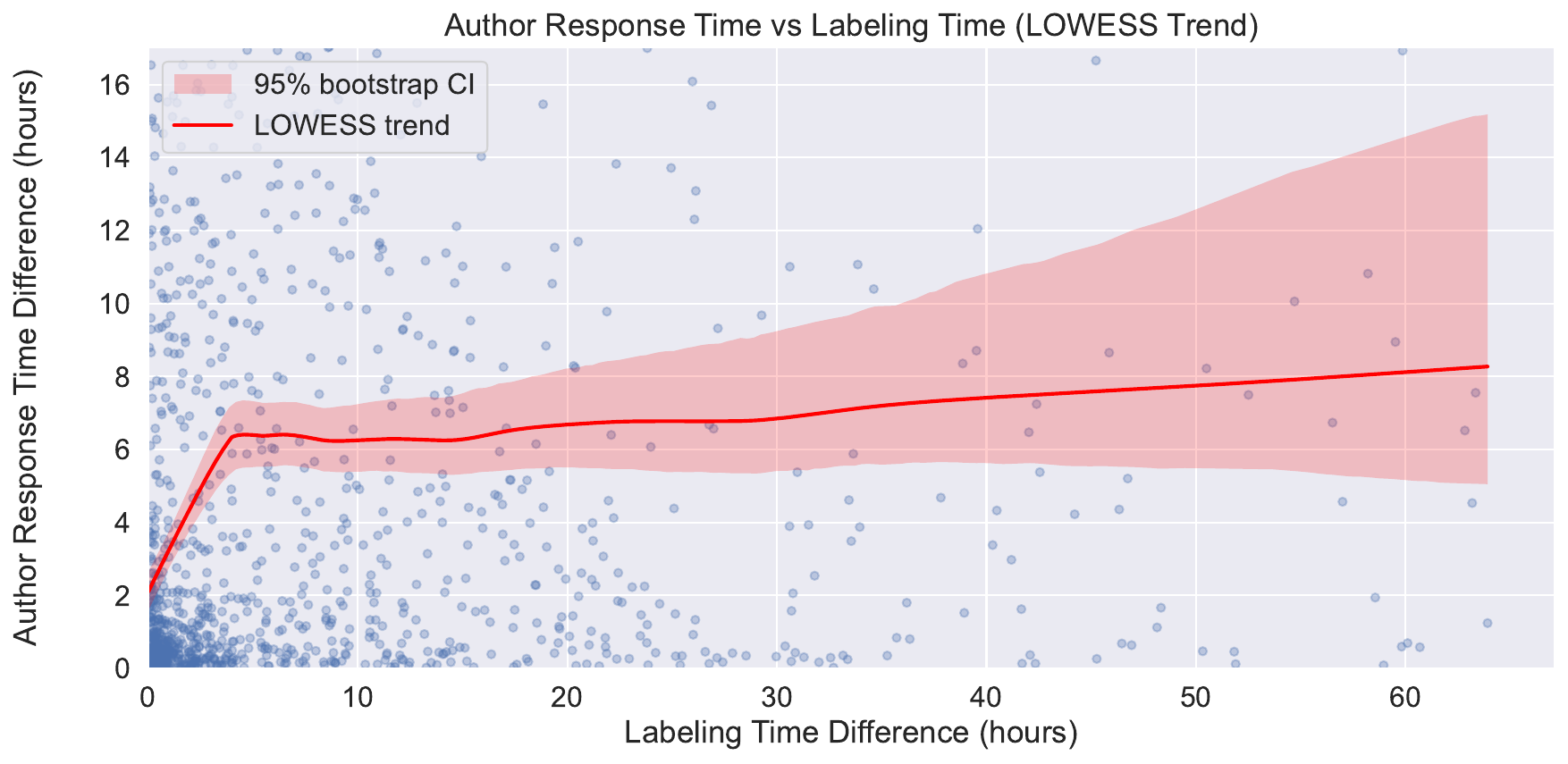}
        \caption{HA: Labeling vs.\ Author Response Time}
        \label{fig:ha_trend_plot}
    \end{subfigure}
    \hfill
    \begin{subfigure}[b]{0.48\linewidth}
        \centering
        \includegraphics[width=\linewidth]{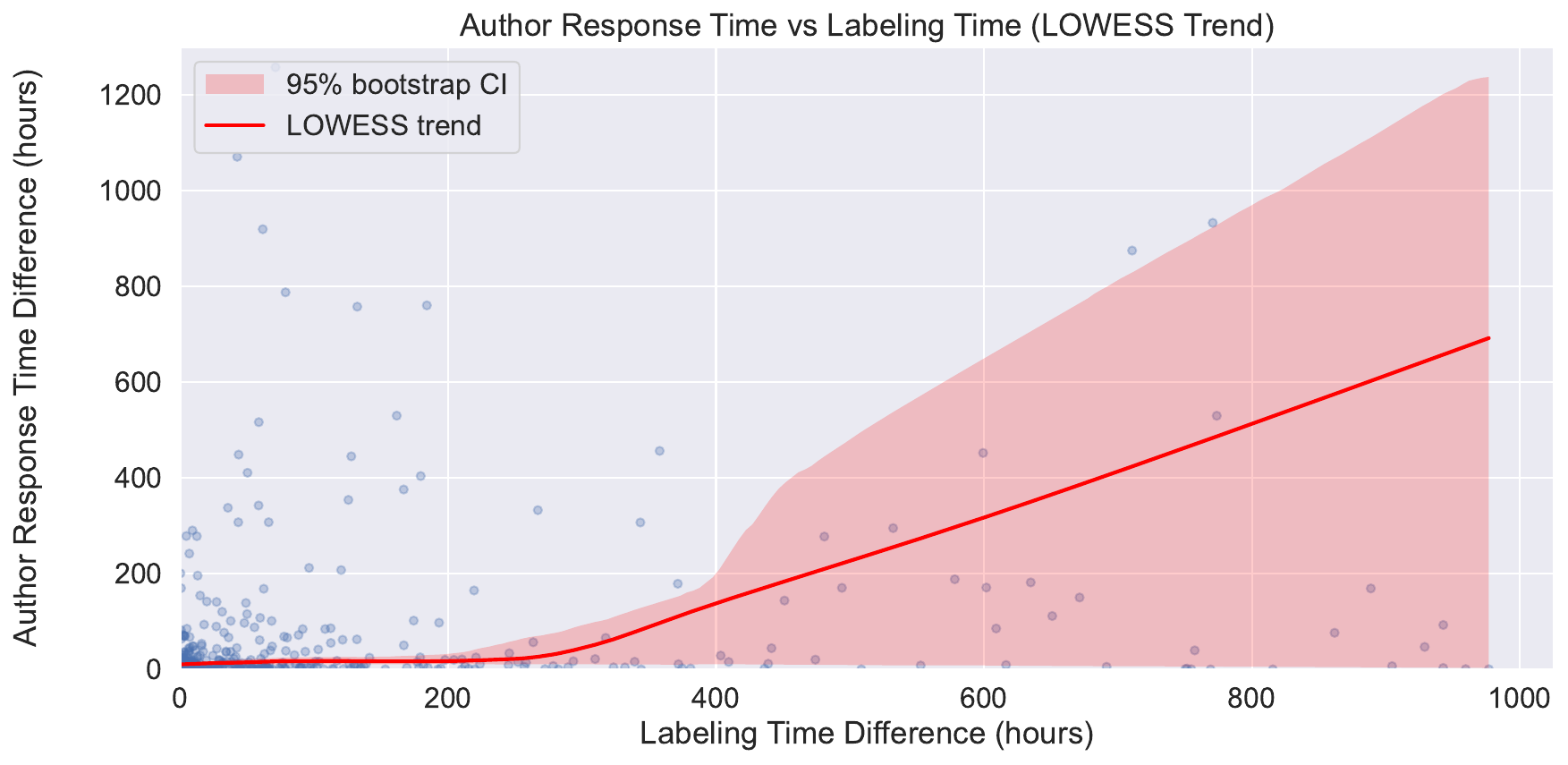}
        \caption{Brave: Labeling vs.\ Author Response Time}
        \label{fig:brave_trend_plot}
    \end{subfigure}

    \caption{Empirical analysis results for the Home Assistant 
    and Brave repositories.}
    \label{fig:empirical_analysis}
\end{figure}
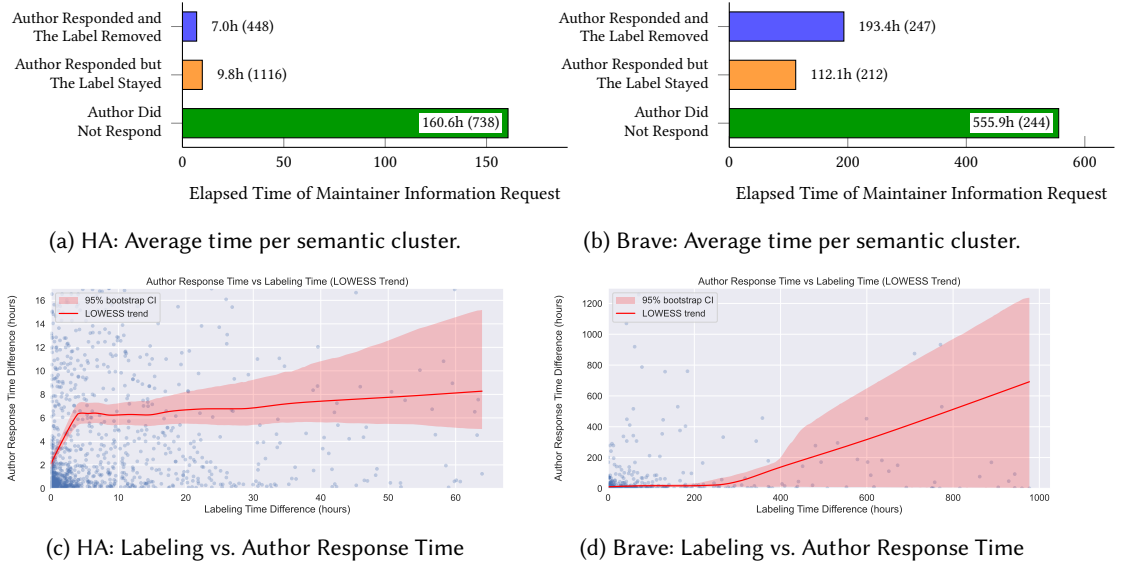


\paragraph{\textbf{\textit{Response Rate Analysis.}}}
\Cref{fig:ha_time_plot} and \Cref{fig:brave_time_plot} show that, in both repositories, information requests were issued substantially later for issues that received no author response than for the two groups in which the author responded. Mann--Whitney U tests confirm that the information request time distributions of both author-replied groups differ significantly from the no-response group ($p < .001$ in both repositories), which suggests that issuing information requests earlier is consistently associated with obtaining an author response.

Among issues in which the author did respond, the information request time differed significantly between the label-removed and label-retained groups in Home Assistant ($p = .001$) but not in Brave (Mann--Whitney U, $p = .279$). The relationship between the timing of information requests and the removal of the \textit{more-information} label is therefore repository-dependent; label removal also depends on repository-specific maintainer practices and the bug triage culture of the repository.

\paragraph{\textbf{\textit{Author Response Time Trend.}}} In both projects, author response time increases with the delay of the maintainer's information request (\Cref{fig:ha_trend_plot,fig:brave_trend_plot}). For Home Assistant, most of this increase occurs within the first few hours, as the smoothed response time rises from $\sim$2\,h for immediate information requests to $\sim$6\,h at a 4-hour delay, and then grows more slowly to $\sim$8\,h. Information requests made within one hour receive a median reply in 0.52\,h, compared with 9.76\,h for those made after four hours (Mann--Whitney $p < .001$, Cliff's $\delta = .514$, large).

For Brave, the information request delays in the trend span a much wider range (up to $\sim$40 days), and the smoothed response time stays around $\sim$17\,h for the first $\sim$10 days before rising at longer delays, where the wide 95\% confidence band reflects sparse data and calls for caution. Because the smoother operates over this wide range, it does not resolve differences within the first hours; the group comparison does; as information requests made within one hour receive a median reply in 1.45\,h, compared with 13.49\,h for those made after four hours (Mann--Whitney $p < .001$, Cliff's $\delta = .377$, medium). Since our framework can generate follow-up questions at submission time, these findings motivate automated, immediate information requests as a way to shorten the feedback loop between maintainers and authors.


\subsection{Question Generator Performance}
\label{sec:rq2_results}

\begin{figure}[t]
\centering
\definecolor{coverageFull}{HTML}{0072B2}
\definecolor{coveragePartial}{HTML}{56B4E9}
\definecolor{coverageNone}{HTML}{C8C8C8}
\pgfplotsset{coveragepanel/.style={
  scale only axis, width=0.72\linewidth, height=3.4cm,
  xmin=-0.6, xmax=2.6, ymin=0,
  axis lines*=left, tick align=inside,
  xtick={0,1,2}, xticklabels={{Gemini 3\\Flash\\Preview},{Qwen 3.8\\Flash},{DeepSeek\\V4.1\\Flash}}, xtick style={draw=none},
  xticklabel style={font=\tiny, align=center}, yticklabel style={font=\scriptsize},
  ylabel style={font=\scriptsize\bfseries},
  area legend,
  legend style={font=\tiny, draw=none, fill=none, legend columns=-1, column sep=3pt, at={(0.5,1.03)}, anchor=south},
}}
\begin{subfigure}[b]{0.49\textwidth}
\centering
\begin{tikzpicture}
\begin{axis}[
    xbar stacked,
    width=\linewidth,
    height=4.0cm,
    xmin=0, xmax=122,
    ymin=-0.5, ymax=2.5,
    xtick={0,25,50,75,100},
    ytick={0,1,2},
    y dir=reverse,
    yticklabels={
        {Gemini 3\\Flash Preview},
        {Qwen 3.8 Flash},
        {DeepSeek\\V4.1 Flash}
    },
    tick label style={font=\tiny},
    yticklabel style={font=\tiny,align=right},
    xlabel={Share of positive reports (\%)},
    xlabel style={font=\scriptsize},
    bar width=10pt,
    axis lines*=left,
    legend style={
        at={(0.5,1.03)},
        anchor=south,
        legend columns=3,
        font=\tiny,
        draw=none,
        fill=none,
        column sep=3pt
    }
]
    \addplot[fill=coverageFull,draw=coverageFull!75!black]
        coordinates {
            (33.724,0) (25.059,1) (39.344,2)
        };
    \addlegendentry{Fully}

    \addplot[fill=coveragePartial,draw=coveragePartial!75!black]
        coordinates {
            (36.300,0) (34.895,1) (34.660,2)
        };
    \addlegendentry{Partially}

    \addplot[fill=coverageNone,draw=coverageNone!75!black]
        coordinates {
            (29.977,0) (40.047,1) (25.995,2)
        };
    \addlegendentry{Not covered}

    \coordinate (full0) at (axis cs:16.862,0);
    \coordinate (part0) at (axis cs:51.874,0);
    \coordinate (none0) at (axis cs:85.012,0);

    \coordinate (full1) at (axis cs:12.529,1);
    \coordinate (part1) at (axis cs:42.506,1);
    \coordinate (none1) at (axis cs:79.977,1);

    \coordinate (full2) at (axis cs:19.672,2);
    \coordinate (part2) at (axis cs:56.674,2);
    \coordinate (none2) at (axis cs:87.002,2);

    \coordinate (pooled0) at (axis cs:102,0);
    \coordinate (pooled1) at (axis cs:102,1);
    \coordinate (pooled2) at (axis cs:102,2);
\end{axis}

\node[font=\tiny,text=white] at (full0) {144};
\node[font=\tiny] at (part0) {155};
\node[font=\tiny] at (none0) {128};

\node[font=\tiny,text=white] at (full1) {107};
\node[font=\tiny] at (part1) {149};
\node[font=\tiny] at (none1) {171};

\node[font=\tiny,text=white] at (full2) {168};
\node[font=\tiny] at (part2) {148};
\node[font=\tiny] at (none2) {111};

\node[font=\tiny,anchor=west] at (pooled0) {43.2\%};
\node[font=\tiny,anchor=west] at (pooled1) {35.4\%};
\node[font=\tiny,anchor=west] at (pooled2) {51.8\%};
\end{tikzpicture}
\caption{Positive reports by coverage.}
\label{fig:rq2_coverage_buckets}
\end{subfigure}\hfill
\begin{subfigure}[b]{0.49\textwidth}
\centering
\begin{tikzpicture}
\begin{axis}[
    xbar,
    width=\linewidth,
    height=4.0cm,
    xmin=0, xmax=108,
    ymin=-0.5, ymax=2.5,
    xtick={0,25,50,75,100},
    ytick={0,1,2},
    y dir=reverse,
    yticklabels={
        {Gemini 3\\Flash Preview},
        {Qwen 3.8 Flash},
        {DeepSeek\\V4.1 Flash}
    },
    tick label style={font=\tiny},
    yticklabel style={font=\tiny,align=right},
    xlabel={Ground-truth questions covered (\%)},
    xlabel style={font=\scriptsize},
    axis lines*=left,
    point meta=explicit symbolic,
    nodes near coords,
    every node near coord/.append style={
        font=\tiny,
        anchor=west,
        xshift=1pt,
        inner sep=1pt,
        align=left
    },
    legend style={
        at={(0.5,1.03)},
        anchor=south,
        legend columns=2,
        font=\tiny,
        draw=none,
        fill=none
    }
]
    \addplot[
        fill=specificityGeneric,
        draw=specificityGeneric!75!black,
        bar width=8pt,
        bar shift=4pt
    ] coordinates {
        (63.06,0) [{63.1\% (227)}]
        (50.83,1) [{50.8\% (183)}]
        (63.89,2) [{63.9\% (230)}]
    };
    \addlegendentry{Generic (360)}

    \addplot[
        fill=specificityDetailed,
        draw=specificityDetailed!75!black,
        bar width=8pt,
        bar shift=-4pt
    ] coordinates {
        (33.24,0) [{33.2\% (237)}]
        (27.63,1) [{27.6\% (197)}]
        (45.72,2) [{45.7\% (326)}]
    };
    \addlegendentry{Detailed (713)}
\end{axis}
\end{tikzpicture}
\caption{Ground-truth questions covered, by specificity.}
\label{fig:rq2_specificity}
\end{subfigure}

\caption{Coverage of the maintainers' follow-up questions on the 427 positive reports. Panel (a) shows the share of reports whose ground-truth questions were fully, partially and not covered. Panel (b) shows the share of the 360 generic and the 713 detailed ground-truth questions that were covered.}
\label{fig:rq2_coverage}
\end{figure}
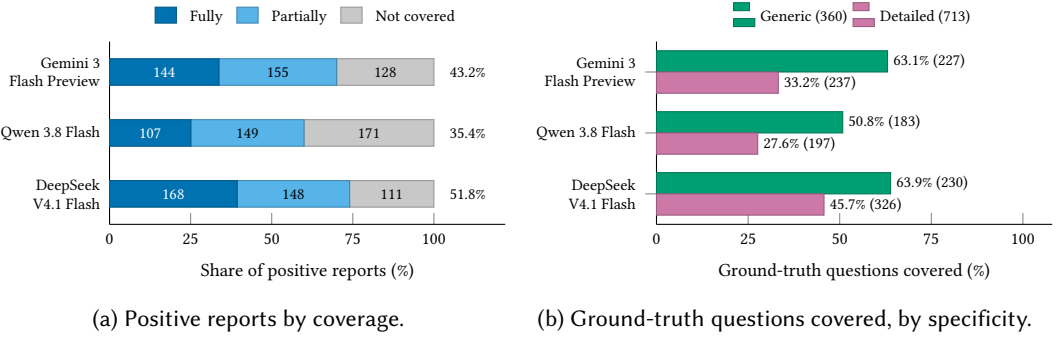

\Cref{fig:rq2_coverage_buckets} shows how completely the generated questions covered the maintainers' follow-up questions. DeepSeek performed best across all coverage analyses, covering 556 of the 1,073 ground-truth questions (51.8\%), fully covering 168 of the 427 positive reports (39.3\%), and leaving only 111 reports (26.0\%) uncovered. In the partially covered reports, the models covered 40.6\% (Qwen) to 47.4\% (DeepSeek) of the ground-truth questions. The models ranked in the same order in both repositories. DeepSeek, Gemini and Qwen covered 49.2\%, 41.7\%, and 32.2\% of the 295 ground-truth questions of Brave, and 52.8\%, 43.8\%, and 36.6\% of the 778 ground-truth questions of Home Assistant.

Coverage generally decreased as the number of ground-truth questions in a report grew. The pooled coverage fell from 50.0\%-61.5\% for the 156 reports with a single ground-truth question, to 22.7\%-30.3\% for the 34 reports with six to ten ground-truth questions. These 34 reports contain 238 of the 1,073 ground-truth questions (22.2\%), so they lower the pooled coverage of every model.

\begin{table*}[t]
\centering
\footnotesize
\caption{Ranking metrics of the generated questions in iterations 1 and 2. The best value per row and metric is in bold.}

\label{tab:ir_metrics}
\setlength{\tabcolsep}{4pt}
\begin{DIFnomarkup}
\begin{tabular}{cc cccc @{\hskip 0.15in} cccc @{\hskip 0.15in} cccc}
\toprule
\multirow{2}{*}{\textbf{Iter}} & \multirow{2}{*}{\textbf{k}} & \multicolumn{4}{c}{\textbf{Gemini 3 Flash Preview}} & \multicolumn{4}{c}{\textbf{Qwen 3.8 Flash}} & \multicolumn{4}{c}{\textbf{DeepSeek V4.1 Flash}} \\
\cmidrule(lr){3-6} \cmidrule(lr){7-10} \cmidrule(lr){11-14}
& & \textbf{P} & \textbf{R} & \textbf{NDCG} & \textbf{MRR} & \textbf{P} & \textbf{R} & \textbf{NDCG} & \textbf{MRR} & \textbf{P} & \textbf{R} & \textbf{NDCG} & \textbf{MRR} \\
\midrule
\multirow{3}{*}{1}
& @1 & \textbf{.520} & \textbf{.310} & \textbf{.520} & \textbf{.520} & .424 & .241 & .424 & .424 & .389 & .222 & .389 & .389 \\
& @3 & \textbf{.299} & \textbf{.473} & \textbf{.472} & \textbf{.599} & .249 & .391 & .390 & .513 & .283 & .460 & .405 & .512 \\
& @5 & .183 & .487 & \textbf{.454} & \textbf{.602} & .155 & .402 & .375 & .514 & \textbf{.203} & \textbf{.524} & .415 & .526 \\
\midrule
\multirow{3}{*}{2}
& @1 & .071 & .019 & .071 & .071 & \textbf{.105} & \textbf{.034} & \textbf{.105} & \textbf{.105} & .088 & .026 & .088 & .088 \\
& @3 & .042 & .036 & .049 & .094 & \textbf{.052} & \textbf{.048} & \textbf{.067} & \textbf{.125} & .051 & .043 & .059 & .110 \\
& @5 & .025 & .036 & .042 & .094 & .031 & \textbf{.048} & \textbf{.060} & \textbf{.125} & \textbf{.033} & .047 & .053 & .113 \\
\bottomrule
\end{tabular}
\end{DIFnomarkup}
\end{table*}

\Cref{tab:ir_metrics} shows how well the models ranked their questions. In iteration 1, Gemini 3 Flash Preview ranked its questions best. Its first question matched a ground-truth question in 52.0\% of the reports (P@1 = .520), compared with 42.4\% for Qwen 3.8 Flash and 38.9\% for DeepSeek V4.1 Flash, and it reached the highest NDCG and MRR for every $k$, with an MRR@5 of .602. DeepSeek V4.1 Flash surpassed it only in precision and recall at $k = 5$, with the highest R@5, .524, compared with .487 for Gemini 3 Flash Preview. This difference follows from the number of questions that the models asked. In iteration 1, DeepSeek V4.1 Flash asked 4.40 questions per report on average, compared with 2.89 for Gemini 3 Flash Preview and 2.66 for Qwen 3.8 Flash. Gemini 3 Flash Preview asked four or more questions on only 96 of the 423 reports on which it asked any, and Qwen 3.8 Flash on 58 of 413, compared with 331 of 427 for DeepSeek V4.1 Flash. The recall of Gemini 3 Flash Preview and Qwen 3.8 Flash therefore changed little beyond $k = 3$, and the questions they did not ask lowered their precision at higher $k$, as they count as incorrect. The higher coverage of DeepSeek V4.1 Flash thus comes from asking more questions, not from ranking them better.

In iteration 2, all metrics dropped. P@1 lay between .071 and .105, and R@5 was at most .048. Two properties of the setup contribute to this drop. Iteration 2 can match only the ground-truth questions that iteration 1 did not already match and answer, while recall still divides by all ground-truth questions of the report. Moreover, the Answer Evaluator asks only about gaps that arise from the answers of iteration 1. Qwen 3.8 Flash scored highest in iteration 2, but on the fewest reports as it generated questions in iteration 2 on 172 reports, compared with 253 for Gemini 3 Flash Preview and 272 for DeepSeek V4.1 Flash.

The first two iterations produced nearly all of the coverage. In iteration 1, the questions covered 430 of the 1,073 ground-truth questions for Gemini 3 Flash Preview, 356 for Qwen 3.8 Flash, and 514 for DeepSeek V4.1 Flash. 
Between 216 and 274 reports per model reached iteration 2, and these reports had only 320 to 353 ground-truth questions available, of which iteration 2 covered 23 to 37. At most 28 reports per model reached iteration 3, at most 4 reached iteration 4, and a single report, with Gemini 3 Flash Preview, reached iteration 5, where no ground-truth question was available. Most fully covered reports therefore became fully covered in iteration 1 with 135 of 144 for Gemini 3 Flash Preview, 103 of 107 for Qwen 3.8 Flash, and 156 of 168 for DeepSeek V4.1 Flash. Nevertheless, \tool~often continued to ask after every ground-truth question had been matched; as a later iteration asked at least one further question in 121 of the 144 fully covered reports of Gemini 3 Flash Preview (84.0\%), in 70 of the 107 of Qwen 3.8 Flash (65.4\%), and in 142 of the 168 of DeepSeek V4.1 Flash (84.5\%). These questions go beyond what the maintainers asked.

\Cref{fig:rq2_specificity} separates generic from detailed ground-truth questions. All three models covered generic questions, whose information the issue template of the project already requests, more often than detailed ones: 63.1\% compared with 33.2\% for Gemini 3 Flash Preview, 50.8\% compared with 27.6\% for Qwen 3.8 Flash, and 63.9\% compared with 45.7\% for DeepSeek V4.1 Flash. This gap held for every model in both repositories. Gemini 3 Flash Preview and DeepSeek V4.1 Flash covered almost the same share of the generic questions, so the higher overall coverage of DeepSeek V4.1 Flash came from the detailed questions, of which it covered 326, compared with 237 for Gemini 3 Flash Preview.

In summary, without access to the maintainers' questions, the generated questions covered 35.4\% to 51.8\% of the 1,073 follow-up questions that maintainers asked, and with DeepSeek V4.1 Flash at least one of them in 74.0\% of the reports. Most of this coverage was reached in the first iteration, before the reporter answered any question. Coverage was highest for reports with a single follow-up question (50.0\% to 61.5\%) and for generic questions (50.8\% to 63.9\%), and the models also covered 27.6\% to 45.7\% of the detailed questions, which the issue template does not anticipate.

\subsection{Bug Report Evaluator Performance}
\label{sec:rq3_results}

\begin{table}[t]
\centering
\footnotesize
\caption{Completeness verdicts at the first iteration on the 427 positive and 429 negative reports under each severity threshold. A report is insufficient when at least one listed field has the threshold severity or a higher one. The threshold \textit{low} is the verdict of \tool; \textit{medium} and \textit{high} are post-hoc analyses.}
\label{tab:checker_first_iteration}

\begin{tabular}{@{}ll rrr@{}}
\toprule
\textbf{Severity threshold} & \textbf{Model} & \textbf{\begin{tabular}[b]{@{}c@{}}Positives judged\\INSUFFICIENT\end{tabular}} & \textbf{\begin{tabular}[b]{@{}c@{}}Negatives judged\\SUFFICIENT\end{tabular}} & \textbf{Macro-F1} \\
\midrule
\multirow{3}{*}{LOW (\tool)} & Gemini 3 Flash Preview & 423 (99.1\%) & 72 (16.8\%) & .493 \\
 & Qwen 3.8 Flash & 419 (98.1\%) & 39 (9.1\%) & .421 \\
 & DeepSeek V4.1 Flash & 427 (100.0\%) & 1 (0.2\%) & .335 \\
\midrule
\multirow{3}{*}{MEDIUM} & Gemini 3 Flash Preview & 419 (98.1\%) & 118 (27.5\%) & .575 \\
 & Qwen 3.8 Flash & 412 (96.5\%) & 114 (26.6\%) & .561 \\
 & DeepSeek V4.1 Flash & 427 (100.0\%) & 13 (3.0\%) & .366 \\
\midrule
\multirow{3}{*}{HIGH} & Gemini 3 Flash Preview & 382 (89.5\%) & 224 (52.2\%) & .698 \\
 & Qwen 3.8 Flash & 344 (80.6\%) & 287 (66.9\%) & .736 \\
 & DeepSeek V4.1 Flash & 372 (87.1\%) & 206 (48.0\%) & .663 \\
\midrule
\multicolumn{2}{@{}l}{Baseline (all INSUFFICIENT)} & 427 (100.0\%) & 0 (0.0\%) & .333 \\
\multicolumn{2}{@{}l}{Baseline (all SUFFICIENT)} & 0 (0.0\%) & 429 (100.0\%) & .334 \\
\bottomrule
\end{tabular}

\end{table}

\Cref{tab:checker_first_iteration} shows the completeness verdicts at the first iteration. All three models judged almost every positive report insufficient. Gemini 3 Flash Preview judged 423 reports (99.1\%), Qwen 3.8 Flash 419 (98.1\%), and DeepSeek V4.1 Flash all 427 (100.0\%) as insufficient out of 427 positive reports. On the negative reports, Gemini 3 Flash Preview judged 72 of the 429 reports (16.8\%) sufficient, Qwen 3.8 Flash 39 (9.1\%), and DeepSeek V4.1 Flash 1 (0.2\%). Recognizing complete reports was therefore the harder of the two decisions for all three models.

As a baseline, we use two verdicts. One verdict labels every report insufficient, and the other verdict labels every report sufficient. Gemini 3 Flash Preview reached a macro-F1 of .493 and improves substantially on the baseline, by .160. Qwen 3.8 Flash reached .421, which exceeds the baseline by .088. DeepSeek V4.1 Flash reached .335, close to the baseline.

\textbf{\textit{Severity thresholds.}} 
In the post-hoc analysis, we raised the severity ($\sigma_{k}$) threshold of the verdict (\textit{medium} and \textit{high} in \Cref{tab:checker_first_iteration}). With the threshold \textit{medium}, which judges a report whose fields are all low as sufficient, the models recognized more complete reports. Gemini 3 Flash Preview judged 118 of the 429 negative reports (27.5\%) sufficient, Qwen 3.8 Flash 114 (26.6\%), and DeepSeek V4.1 Flash 13 (3.0\%), while still judging 96.5\% to 100.0\% of the positive reports insufficient. Macro-F1 rose to 57.5\%, 56.1\%, and 36.6\%, respectively. With the threshold \textit{high}, the models recognized many more complete reports, 224 (52.2\%), 287 (66.9\%), and 206 (48.0\%), in exchange for judging fewer positive reports insufficient; as 382 (89.5\%), 344 (80.6\%), and 372 (87.1\%). Macro-F1 rose to 69.8\%, 73.6\%, and 66.3\%, and Qwen 3.8 Flash reached the highest score. The threshold \textit{high} raised macro-F1 in both repositories, to 59.7\%-72.6\% for Brave and 69.3\%-74.0\% for Home Assistant.

Because a report judged sufficient receives no questions, a higher threshold also affects RQ2. A report judged insufficient runs exactly as logged, and a report judged sufficient loses its coverage, so the effect follows from the same runs. With the threshold \textit{medium}, the pooled coverage of the 1,073 ground-truth questions stayed almost the same (43.1\%, 35.4\%, and 51.8\%). With the threshold \textit{high}, 
the pooled coverage decreased to 41.2\%, 32.1\%, and 48.1\%. The severities that the models assign therefore separate complete from incomplete reports better than the mere presence of a field. 

\subsection{Token Usage}
\label{sec:token_usage}

We recorded every LLM call of \tool~with Langfuse\footnote{\url{https://langfuse.com}} and summed the tokens of each report over the calls of the run that produced its result; the User Simulator is not included, and calls that failed without a response and were retried used no tokens. Over all 856 reports, a report required on average 8,462 (Qwen 3.8 Flash) to 10,457 (DeepSeek V4.1 Flash) input tokens and 516 (Gemini 3 Flash Preview) to 725 (DeepSeek V4.1 Flash) output tokens. A positive report required 11,969 to 16,108 input tokens, 2.4 to 3.3 times as many as a negative report (4,832 to 5,119), because \tool~continued with further iterations on it, as 1.60 to 1.87 calls per positive report, compared with 1.00 to 1.03 per negative report. DeepSeek V4.1 Flash, which asked the most questions, used the most tokens on the positive reports. Each model counts tokens with its own tokenizer, so the counts are not directly comparable across models.

\section{Discussion}

\paragraph{\textbf{\textit{Implications for Researchers.}}} Future research should focus on enhancing the pipeline through advanced context engineering. Such context could especially help with detailed questions, which the issue template does not anticipate and which all three models covered less often than generic ones (Section~\ref{sec:rq2_results}). Retrieval-Augmented Generation (RAG) could supply domain-specific sources such as project wikis, source code repositories, and historical bug reports, and agentic workflows with web search could retrieve external guidelines or tutorials that help authors answer complex follow-up questions. Beyond the context that the models receive, researchers could examine whether enabling reasoning, which our setup disabled for all three models, changes the completeness verdict or the generated questions. To establish the generalizability of the framework, experiments should also include proprietary, industrial software projects with other issue-reporting standards. On the evaluation side, future work should complement quantitative metrics with qualitative assessments of the generated questions, such as their clarity, relevance, and perceived usefulness. A field study could extend this evaluation by deploying \tool~in a live issue tracker, where real authors, rather than the User Simulator, answer the generated questions. Such a study would show how often and how quickly authors answer questions asked at submission time, and whether their answers reduce the time that maintainers need to resolve a report. Finally, the follow-up questions that depend on earlier answers could be evaluated separately. The iterations after the first covered only 24 to 42 ground-truth questions, and a separate evaluation would show whether this reflects the Answer Evaluator or the small number of such questions in the ground truth.



\paragraph{\textbf{\textit{Implications for Practitioners.}}} The evaluated models judged most complete reports insufficient, so a deployment that sends generated questions to the author would also question many authors whose reports need no further information. To overcome this problem, the practitioners can configure their completeness threshold based on the missing gap severity, as in our post-hoc analysis in RQ3. The choice of model also matters; as in our evaluation, Gemini 3 Flash Preview recognized more complete reports than the other models and most often ranked a matching question first, whereas DeepSeek V4.1 Flash asked the most questions in the first iteration and covered the most ground-truth questions (Section~\ref{sec:rq2_results}). \tool~accesses its models through OpenRouter, so practitioners can replace the evaluated models by changing the model identifier. Practitioners may select the more cost-performance model for the pipeline based on the given performance analysis on Section~\ref{sec:rq2_results} to~\ref{sec:rq3_results}; and token usage analysis on Section~\ref{sec:token_usage}. Moreover, our pipeline requires no additional project-specific information injection; so that practitioners can easily configure the pipeline into their own workflows. Unlike our study (Section~\ref{sec:experimental_setup}), a deployment may also use reasoning or the default sampling temperature, although reasoning may delay the questions. \tool~needs only the submitted report and the author's answers as input, so it can be integrated into the platform where the project collects bug reports. For example on GitHub, a bot triggered by a new issue can post the questions as a comment. Practitioners can further configure the maximum number of iterations and the number of questions asked per iteration, trading coverage of missing information against the number of questions the author receives.




\section{Threats to Validity}


\paragraph{\textbf{\textit{Internal Validity.}}}
\tool~is evaluated using a User Simulator that answers each generated question by matching it to the report's ground-truth questions. The simulator may either incorrectly match a question, providing an answer it should not receive, or miss a valid match, withholding an answer it should receive. Because these answers become part of the context evaluated in subsequent iterations, both errors can affect later generated questions. We measured simulator agreement with human annotators on a stratified sample of 59 bug reports, with agreements of 0.82 precision, 0.85 recall and 0.84 F1 score; downstream results should therefore be interpreted in light of this agreement. The experiments also rely on black-box LLMs whose outputs may vary between runs or change as services are updated. We set temperature to zero to reduce run-to-run variation.

\paragraph{\textbf{\textit{External Validity.}}}
Our results may not generalize to all software projects because we evaluate \tool~on only two open-source repositories. To increase diversity, we selected repositories that differ substantially in domain, issue templates, and user base: \textit{home-assistant/core}, a home-automation platform with many third-party device integrations, and \textit{brave/brave-browser}, a web browser. Thus, the evaluation is not restricted to the reporting conventions of a single project. We aim to evaluate \tool~on closed-source or industrial issue trackers, where reporting and maintenance practices may differ, as future work.

\paragraph{\textbf{\textit{Construct Validity.}}}
Our evaluation treats correspondence with questions actually asked by maintainers as evidence of question quality. A generated question may nevertheless be useful even if no matching ground-truth question exists, causing the framework to underestimate its quality. Conversely, maintainer-asked questions provide a practical signal of which information maintainers considered necessary within the workflow of each repository. We therefore view ground-truth question matching as a complementary measure grounded in real maintenance behavior. Moreover, the negative set was not manually verified and may include incomplete reports. To reduce this noise, we retained only reports with no author comments or edits and an associated merged fix, suggesting that maintainers could resolve the issue without further input from the author. Finally, some evaluated bug reports were resolved before the knowledge cutoff of the tested models and may therefore have appeared in their training data.

\section{Conclusion}

In this study, we introduced \tool, a semi-automated framework that assesses whether a bug report is complete and generates follow-up questions for the information it lacks, and repeats the process for the author-replied answers to the generated questions; in order to enhance author engagement and the bug triage process with newly introduced information. Our empirical study on \textit{Home Assistant} and \textit{Brave Browser} showed that maintainers' earlier information requests are significantly associated with more frequent and faster author responses. We introduced a novel dataset of 429 incomplete bug reports with 1,081 real maintainer-asked questions with 362 generic and 719 detailed questions, mapped to their author-replied answers. Across three models, \tool~covered 380 to 556 of the 1,073 maintainer questions (35.4\%-51.8\%), mostly in the first two iterations, and by specificity, 183 to 230 generic questions (50.8\%-63.9\%) and 197 to 326 detailed questions (27.6\%-45.7\%) that required information beyond the issue template, which can be automated with \tool. Initially, all models marked at least 98.1\% of incomplete reports insufficient but at most 16.8\% of complete reports sufficient. Requiring at least one high severity gap retained at least 80.6\% of incomplete reports as insufficient and raised recognition of complete reports to as much as 66.9\%. The best macro-F1 rose from 49.3\% (Gemini 3 Flash Preview) to 73.6\% (Qwen 3.8 Flash), versus 33.4\% for the always-sufficient baseline. These results show that \tool~can recover detailed questions beyond those prompted by issue templates, while the completeness threshold remains important for deciding when to stop asking. Future work will focus on integrating context engineering via RAG and validating the pipeline on industrial software ecosystems. 

\section{Data Availability}
\label{sec:data_availability}

The dataset, source code, utilized prompts and the complete replication package are available to support reproducibility at \url{https://figshare.com/s/25a43a746faa7cee311d}.

\section{Acknowledgments}
\label{sec: acknowledgement}
This work was supported by TÜBİTAK (The Scientific and Technological Research Council of Turkey) under the 1001 Scientific and Technological Research Projects Funding Program, Project No. 125E371. The authors gratefully acknowledge TÜBİTAK for its support.

\bibliographystyle{ACM-Reference-Format}
\bibliography{bibfile}

\end{document}